\documentclass[aps,prc,twocolumn,tightenlines,floatfix,showpacs]{revtex4}

\usepackage[dvips]{graphicx}

\usepackage{dcolumn}     

\renewcommand{\case}{\frac}

\begin{document}
 
{

\title{Quantum Monte Carlo calculations of excited states in $A = 6-8$ nuclei}

\author{Steven C. Pieper\cite{scp} and R. B. Wiringa\cite{rbw}}

\affiliation{Physics Division, Argonne National Laboratory, 
             Argonne, Illinois 60439}

\author{J. Carlson\cite{jc}}

\affiliation{Theoretical Division, Los Alamos National Laboratory,
             Los Alamos, New Mexico 87545}

\date{\today}
 
\begin{abstract}
A variational Monte Carlo method is used to generate sets of orthogonal
trial functions, $\Psi_T(J^\pi,T)$, for given quantum numbers in various 
light p-shell nuclei.
These $\Psi_T$ are then used as input to Green's function Monte Carlo
calculations of first, second, and higher excited $(J^\pi,T)$ states.
Realistic two- and three-nucleon interactions are used.
We find that if the physical excited state is reasonably narrow, the
GFMC energy converges to a stable result.
With the combined Argonne v$_{18}$ two-nucleon and Illinois-2 three-nucleon
interactions, the results for many second and higher states in 
$A$ = 6--8 nuclei are close to the experimental values.
\end{abstract}
 

\maketitle

}

\section {Introduction}

Quantum Monte Carlo methods have proved to be very accurate and powerful
tools for studying the structure of light nuclei with realistic two- and
three-nucleon interactions.
In a series of papers, we have calculated about 40 ground and low-lying
excited state energies of different $(J^\pi,T)$ quantum numbers in
$A$ = 6--10 nuclei with an accuracy of $\sim$ 1--2\%
\cite{PPCW95,PPCPW97,WPCP00,PPWC01,PVW02}.
The first step is a variational Monte Carlo (VMC) calculation to find an
optimal trial function, $\Psi_T(J^\pi,T)$ for a given state.
The $\Psi_T$, which is antisymmetric by explicit construction, is then used 
as input to a Green's function Monte Carlo (GFMC) algorithm, which projects 
out the lowest-energy eigenstate by a propagation in imaginary time, $\tau$.
The algorithm preserves the quantum numbers of $\Psi_T$, although 
it may introduce symmetric noise into the propagated $\Psi(\tau)$.
In principle, $\Psi(\tau)$ for large $\tau$ approaches the exact 
lowest-energy eigenstate with the specified quantum numbers.

VMC calculations of second and higher excited $(J^\pi,T)$ states have also
been made, because a major step in the preparation of an optimal $\Psi_T$
is a diagonalization in the small single-particle basis of different possible 
spatial-symmetry states \cite{PPCPW97,WPCP00,PVW02}.
GFMC calculations of second or higher $(J^\pi,T)$ states have not
been attempted previously in nuclear physics, under the expectation that any 
small contamination of the excited trial state by the true 
first state would drive the calculated energy below its proper value.
However, as is described in this paper, we have found that it is possible
to obtain useful GFMC predictions of multiple states with the same
quantum numbers.

In this paper we report such GFMC calculations of second and higher $(J^\pi,T)$ 
states in light ($A$ = 6--8) p-shell nuclei.
VMC calculations of these states were reported previously
\cite{PPCPW97,WPCP00}, but we have made significant improvements in the trial 
functions since then.
Details of the recent VMC work are discussed in Sec.~II.
For the GFMC calculations, we find that the propagated energy, $E(\tau)$, 
for many states is stable and a useful excitation energy can be extracted.
GFMC calculations for ground states are reviewed in Sec.~III, where we
also describe an improved algorithm for propagation with three-nucleon
interactions that is both faster and more accurate than what we had used
previously.
The GFMC algorithm for higher excited states is described in Sec.~IV,
along with various orthogonality tests.
Numerical results are given in Sec.~V for the Argonne v$_{18}$ (AV18)
and simplified v$_{8^\prime}$ (AV8$^\prime$) two-nucleon ($N\!N$) interactions
\cite{WSS95}, and for AV18 with either the Urbana IX (UIX) or Illinois-2 (IL2)
three-nucleon ($N\!N\!N$) interactions \cite{PPCW95,PPWC01} added.
We find, consistent with our studies of first $(J^\pi,T)$ states, that the
AV18/IL2 Hamiltonian gives a good representation of the experimental
spectrum; 
the RMS deviation from 36 experimental energies with $6\leq A\leq 8$ 
is only 0.60 MeV.
Some concluding remarks are given in Sec.~VI.

\section {VMC trial functions}

The VMC trial function, $\Psi_T(J^\pi;T)$, for a given nucleus, is constructed
from products of two- and three-body correlation operators acting on an
antisymmetric single-particle state with the appropriate quantum numbers.
The correlation operators reflect the influence of the interactions at short
distances, while appropriate boundary conditions are imposed at long range.
The $\Psi_T$ contains variational parameters that are adjusted to
minimize the energy expectation value, $E_T = \langle \Psi_T | H |
\Psi_T \rangle / \langle \Psi_T | \Psi_T \rangle$, which is evaluated by 
Metropolis Monte Carlo integration.

A good variational trial function has the form
\begin{equation}
     |\Psi_T\rangle = \left[1 + \sum_{i<j<k} U^{TNI}_{ijk} \right]
                      \left[ {\mathcal S}\prod_{i<j}(1+U_{ij}) \right]
                      |\Psi_J\rangle \ .
\label{eq:psit}
\end{equation}
The $U_{ij}$ and $U^{TNI}_{ijk}$ are non-commuting two- and three-nucleon
correlation operators, ${\mathcal S}$ indicates a symmetric sum over
all possible orderings, and $\Psi_J$ is the fully antisymmetric Jastrow
wave function.
For s-shell nuclei the latter has the simple form
\begin{equation}
     |\Psi_J\rangle = \left[ \prod_{i<j<k}f^c_{ijk} \right]
                      \left[ \prod_{i<j}f_c(r_{ij}) \right]
                     |\Phi_A(JMTT_{3})\rangle \ .
\label{eq:jastrow}
\end{equation}
Here $f_c(r_{ij})$ and $f^c_{ijk}$ are central two- and three-body correlation
functions and $\Phi_A$ is a Slater determinant in spin-isospin space, e.g.,
for the $\alpha$-particle, $|\Phi_{4}(0 0 0 0) \rangle
= {\mathcal A} |p\uparrow p\downarrow n\uparrow n\downarrow \rangle$.

The correlation operator $U_{ij}$ includes spin, isospin, and tensor terms:
\begin{equation}
     U_{ij} = \sum_{p=2,6} u_p(r_{ij}) O^p_{ij} \ ,
\label{eq:uij}
\end{equation}
where the $O^{p=1,6}_{ij} = [1, {\bf\sigma}_{i}\cdot{\bf\sigma}_{j}, S_{ij} ]
\otimes [1,{\bf\tau}_{i}\cdot{\bf\tau}_{j}]$ are the same static operators
that appear in the $N\!N$ potential.
The functions $f_c(r)$ and $u_p(r)$ are generated by solving a set of six
coupled differential equations with embedded variational
parameters~\cite{W91}: two single-channel equations in $^1$S and $^1$P
waves, and two coupled-channel equations in $^3$S and $^3$P waves.
The $U^{TNI}_{ijk}$ has the spin-isospin structure of the dominant parts
of the $N\!N\!N$ interaction as suggested by perturbation theory.

For p-shell nuclei, the Jastrow wave function, $\Psi_J$, is necessarily
more complicated.
It starts with a sum over independent-particle terms, $\Phi_A$, that have 4
nucleons in an $\alpha$-like core and $(A-4)$ nucleons in p-shell orbitals.
We use $LS$ coupling to obtain the desired $JM$ value of a given state,
as suggested by standard shell-model studies~\cite{CK65}.
We also need to specify the spatial symmetry $[n]$ of the angular
momentum coupling of the p-shell nucleons~\cite{BM69}.
Different possible $LS[n]$ combinations lead to multiple components in the
Jastrow wave function.
This independent-particle basis is again acted on by products of central pair
and triplet correlation functions, but now they depend upon the shells (s or p)
occupied by the particles and on the $LS[n]$ coupling:
\begin{widetext}
\begin{eqnarray}
  |\Psi_{J}\rangle &=& {\mathcal A} \left\{
     \Big[\prod_{i<j<k} f^{c}_{ijk}\Big] \Big[\prod_{i<j \leq 4} f_{ss}(r_{ij})\Big]
     \sum_{LS[n]} \Big( \beta_{LS[n]}
     \Big[\prod_{k \leq 4<l \leq A} f^{LS[n]}_{sp}(r_{kl})\Big] \right.   \nonumber\\
  &\times& \left. \Big[\prod_{4<l<m \leq A} f^{LS[n]}_{pp}(r_{lm})\Big]
    |\Phi_{A}(LS[n]JMTT_{3})_{1234:5\ldots A}\rangle \Big) \right\} \ .
\label{eq:psip}
\end{eqnarray}
The operator ${\mathcal A}$ indicates an antisymmetric sum over all possible
partitions of the $A$ particles into 4 s-shell and $(A-4)$ p-shell ones.
The pair correlation for both particles within the s-shell, $f_{ss}$,
is similar to the $f_c$ of the $\alpha$-particle.
The pair correlations for both particles in the p-shell, $f^{LS[n]}_{pp}$,
and for mixed pairs, $f^{LS[n]}_{sp}$, are similar to $f_{ss}$ at short
distance, but their long-range structure is adjusted to give appropriate
clustering behavior, and they may vary with $LS[n]$.

The single-particle wave functions $\Phi_A$ are given by:
\begin{eqnarray}
  |\Phi_{A}(LS[n]JMTT_{3})_{1234:5\ldots A}\rangle &=&
     |\Phi_{4}(0 0 0 0)_{1234} \Big[\prod_{4 < l\leq A}
     \phi^{LS[n]}_{p}(R_{\alpha l})\Big] \\
 &\times&  \left\{ \Big[ \prod_{4 < l\leq A} Y_{1m_l}(\Omega_{\alpha l}) \Big]_{LM_L[n]}
     \Big[ \prod_{4 < l\leq A} \chi_{l}(\frac{1}{2}m_s) \Big]_{SM_S}
     \right\}_{JM}
     \Big[ \prod_{4 < l\leq A} \nu_{l}(\frac{1}{2}t_3) \Big]_{TT_3}\rangle
     \nonumber \ .
\end{eqnarray}
\end{widetext}
The $\phi^{LS[n]}_{p}(R_{\alpha l})$ are p-wave solutions of a particle
in an effective $\alpha-N$ potential that has Woods-Saxon and Coulomb parts.
They are functions of the distance between the center of mass
of the $\alpha$ core and nucleon $l$, and may vary with $LS[n]$.
The depth, width, and surface thickness of the single-particle potential 
are additional variational parameters of the trial function.
The overall wave function is translationally invariant, so there is no
spurious center of mass motion.

The $\beta_{LS[n]}$ mixing parameters of Eq.~(\ref{eq:psip}) are determined
by a diagonalization procedure, in which matrix elements
\begin{eqnarray}
       E_{T,ij} &=& \langle \Psi_T(\beta_i) | H | \Psi_T(\beta_j) \rangle \ , 
\label{eq:geneigen-e}\\
       N_{T,ij} &=& \langle \Psi_T(\beta_i) | \Psi_T(\beta_j) \rangle \ ,
\label{eq:geneigen-n}
\end{eqnarray}
are evaluated using trial functions $\Psi_T(\beta_i) \equiv \Psi_T(\beta_i=1,
\beta_{j\ne i}=0)$.
Although the $\Phi_A(LS[n]JMTT_3)$ are orthogonal due to spatial symmetry,
the pair and triplet correlations in $\Psi_T$ make the different $LS[n]$
components nonorthogonal, so a generalized eigenvalue routine is necessary
to carry out the diagonalization.

\begin{table}[ht!]
\caption{$\beta_{LS[n]}$ components for $^6$He states, listed in order of
increasing excitation. (L = S or D as appropriate.)}
\begin{tabular}{lrr}
\multicolumn{1} {c}{$(J^{\pi};T)$} & \multicolumn{1} {r}{$^1$L[2]}      &
\multicolumn{1} {r}{$^3$P[11]}     \\
\colrule 
$(0^+;1)$     &    0.974 &   -0.228   \\
$(2^+;1)$     &    0.922 &    0.386   \\
$(2^+;1)$     &   -0.388 &    0.922   \\
$(1^+;1)$     &          &    1.~~~~~ \\  
$(0^+;1)$     &    0.232 &    0.973   \\  
\end{tabular}
\label{tab:beta61}
\end{table}
\vspace*{-.2in}
\begin{table}[ht!]
\caption{$\beta_{LS[n]}$ components in $^6$Li states, listed in order of
increasing excitation.}
\begin{tabular}{lrrr}
\multicolumn{1} {c}{$(J^{\pi};T)$} & \multicolumn{1} {r}{$^3$S[2]}      & 
\multicolumn{1} {r}{$^3$D[2]}      & \multicolumn{1} {r}{$^1$P[11]}     \\
\colrule
$(1^+,0)$     &    0.986 & 0.138    &    0.098   \\
$(3^+;0)$     &          & 1.~~~~~  &            \\
$(2^+;0)$     &          & 1.~~~~~  &            \\
$(1^+,0)$     &   -0.109 & 0.964    &   -0.242   \\
$(1^+,0)$     &   -0.134 & 0.231    &    0.964   \\
\end{tabular}
\label{tab:beta60}
\end{table}
\vspace*{-.2in}
\begin{table}[ht!]
\caption{$\beta_{LS[n]}$ components for $^7$He states, listed in order of
increasing excitation.}
\begin{tabular}{lrrr}
\multicolumn{1} {c}{$(J^{\pi};T)$} & \multicolumn{1} {r}{$^2$P[21]}     & 
\multicolumn{1} {r}{$^2$D[21]}     & \multicolumn{1} {r}{$^4$S[111]}    \\
\colrule
$(\case{3}{2}^-,\case{3}{2})$ &    0.837  &    0.515   &   -0.182   \\
$(\case{1}{2}^-,\case{3}{2})$ &    1.~~~~~&            &            \\
$(\case{5}{2}^-,\case{3}{2})$ &           &    1.~~~~~ &            \\
$(\case{3}{2}^-,\case{3}{2})$ &   -0.392  &    0.805   &    0.445   \\
$(\case{3}{2}^-,\case{3}{2})$ &    0.373  &   -0.317   &    0.872   \\
\end{tabular}
\label{tab:beta73}
\end{table}
\vspace*{-.2in}
\begin{table}[ht!]
\caption{$\beta_{LS[n]}$ components for $^7$Li states, listed in order of
increasing excitation. (L = P or F as appropriate.)}
\begin{tabular}{lrrrrrr}
\multicolumn{1} {c}{$(J^{\pi};T)$} & \multicolumn{1} {r}{$^2$L[3]}      & 
\multicolumn{1} {r}{$^4$P[21]}     & \multicolumn{1} {r}{$^4$D[21]}     &
\multicolumn{1} {r}{$^2$P[21]}     & \multicolumn{1} {r}{$^2$D[21]}     & 
\multicolumn{1} {r}{$^2$S[111]}    \\
\colrule
$(\case{3}{2}^-,\case{1}{2})$ &    0.995  &    0.086  &    0.026
                              &    0.005  &   -0.050  &              \\
$(\case{1}{2}^-,\case{1}{2})$ &    0.988  &   -0.003  &   -0.098
                              &   -0.116  &           &   -0.024     \\
$(\case{7}{2}^-,\case{1}{2})$ &    0.990  &           &    0.138
                              &           &           &              \\
$(\case{5}{2}^-,\case{1}{2})$ &    0.988  &    0.114  &    0.072
                              &           &   -0.079  &              \\
$(\case{5}{2}^-,\case{1}{2})$ &   -0.079  &    0.957  &   -0.058
                              &           &    0.274  &              \\
$(\case{3}{2}^-,\case{1}{2})$ &   -0.084  &    0.952  &    0.250
                              &   -0.087  &    0.132  &              \\
$(\case{1}{2}^-,\case{1}{2})$ &    0.049  &    0.841  &   -0.111
                              &    0.523  &           &   -0.069     \\
$(\case{7}{2}^-,\case{1}{2})$ &   -0.129  &           &    0.992
                              &           &           &              \\
$(\case{5}{2}^-,\case{1}{2})$ &   -0.088  &   -0.009  &    0.968
                              &           &    0.235  &              \\
$(\case{5}{2}^-,\case{1}{2})$ &    0.137  &   -0.263  &   -0.224
                              &           &    0.928  &              \\
\end{tabular}
\label{tab:beta71}
\end{table}
\begin{table}[ht!]
\caption{$\beta_{LS[n]}$ components for $^8$He states, listed in order of
increasing excitation.  (L = S or D as appropriate.)}
\begin{tabular}{lrr}
\multicolumn{1} {c}{$(J^{\pi};T)$} & \multicolumn{1} {r}{$^1$L[22]} & 
\multicolumn{1} {r}{$^3$P[211]}    \\
\colrule
$(0^+;2)$     &   0.794   &  -0.608    \\
$(2^+;2)$     &   0.928   &  -0.373    \\
$(1^+;2)$     &           &   1.~~~~~  \\
$(0^+;2)$     &   0.610   &   0.792    \\
$(2^+;2)$     &   0.377   &   0.926    \\
\end{tabular}
\label{tab:beta82}
\end{table}
\begin{table*}[ht!]
\caption{$\beta_{LS[n]}$ components for $^8$Li states, listed in order of
increasing excitation.  (L = P, D or F as appropriate.)}
\begin{tabular}{lrrrrrrrrr}
\multicolumn{1} {c}{$(J^{\pi};T)$} & \multicolumn{1} {r}{$^3$P[31]}   & 
\multicolumn{1} {r}{$^3$D[31]}     & \multicolumn{1} {r}{$^3$F[31]}   & 
\multicolumn{1} {r}{$^1$L[31]}     & \multicolumn{1} {r}{$^3$S[22]}   & 
\multicolumn{1} {r}{$^3$D[22]}     & \multicolumn{1} {r}{$^5$P[211]}  & 
\multicolumn{1} {r}{$^3$P[211]}    & \multicolumn{1} {r}{$^1$P[211]}  \\
\colrule
$(2^+;1)$     &   0.948   &  -0.248   &  -0.094   &   0.122
              &           &  -0.116   &   0.047   &   0.019   &             \\
$(1^+;1)$     &   0.772   &  -0.299   &           &   0.524
              &  -0.091   &  -0.074   &   0.120   &  -0.112   &  -0.005     \\
$(3^+;1)$     &           &   0.947   &   0.126   &  -0.257
              &           &  -0.061   &  -0.135   &           &             \\
$(0^+;1)$     &   0.995   &           &           &
              &           &           &           &  -0.100   &             \\
$(1^+;1)$     &  -0.588   &  -0.302   &           &   0.733
              &   0.123   &   0.089   &  -0.025   &  -0.031   &  -0.009     \\
$(2^+;1)$     &   0.281   &   0.908   &   0.297   &  -0.038
              &           &   0.040   &  -0.056   &   0.057   &             \\
$(2^+;1)$     &  -0.079   &  -0.128   &   0.533   &   0.823
              &           &  -0.092   &  -0.053   &   0.063   &             \\
$(1^+;1)$     &   0.051   &   0.873   &           &   0.412
              &  -0.221   &   0.065   &  -0.045   &  -0.064   &   0.078     \\
$(3^+;1)$     &           &  -0.153   &   0.945   &  -0.125
              &           &  -0.229   &   0.126   &           &             \\
$(4^+;1)$     &           &           &   1.~~~~~ &
              &           &           &           &           &             \\
\end{tabular}
\label{tab:beta81}
\end{table*}
\begin{table*}[ht!]
\caption{$\beta_{LS[n]}$ components for $^8$Be states, listed in order of
increasing excitation.  (L = S, D or G as appropriate.)}
\begin{tabular}{lrrrrrrrrr}
\multicolumn{1} {c}{$(J^{\pi};T)$} & \multicolumn{1} {r}{$^1$L[4]}    & 
\multicolumn{1} {r}{$^3$P[31]}     & \multicolumn{1} {r}{$^3$D[31]}   & 
\multicolumn{1} {r}{$^3$F[31]}     & \multicolumn{1} {r}{$^5$S[22]}   & 
\multicolumn{1} {r}{$^5$D[22]}     & \multicolumn{1} {r}{$^1$L[22]}   & 
\multicolumn{1} {r}{$^3$P[211]}    \\
\colrule
$(0^+;0)$     &   0.998   &  -0.055   &           &
              &           &  -0.034   &   0.016   &   0.015    \\
$(2^+;0)$     &   0.998   &   0.026   &   0.041   &  -0.020
              &   0.002   &   0.029   &   0.007   &   0.008    \\
$(4^+;0)$     &   0.999   &           &           &   0.052
              &           &  -0.015   &           &            \\
$(2^+;0)$     &  -0.011   &   0.931   &  -0.169   &   0.087
              &   0.284   &  -0.062   &  -0.101   &  -0.055    \\
$(1^+;0)$     &           &   0.957   &   0.237   &
              &           &  -0.165   &           &   0.018    \\
$(1^+;0)$     &           &  -0.249   &   0.959   &
              &           &   0.098   &           &   0.097    \\
$(3^+;0)$     &           &           &   0.936   &   0.287
              &           &   0.207   &           &            \\
$(4^+;0)$     &  -0.018   &           &           &   0.691
              &           &   0.723   &           &            \\
$(2^+;0)$     &  -0.038   &  -0.006   &   0.825   &   0.230
              &   0.499   &   0.050   &   0.111   &   0.039    \\
$(0^+;0)$     &   0.029   &   0.926   &           &
              &           &  -0.311   &   0.181   &  -0.105    \\
$(3^+;0)$     &           &           &  -0.127   &   0.782
              &           &  -0.610   &           &            \\
$(2^+;0)$     &   0.016   &  -0.301   &  -0.376   &  -0.257
              &   0.782   &   0.148   &  -0.011   &  -0.261    \\
\vspace*{.7in}
\end{tabular}
\label{tab:beta80}
\end{table*}

The results for the $\beta_{LS[n]}$ coefficients are shown in 
Tables~\ref{tab:beta61}--\ref{tab:beta80}.
Tables~\ref{tab:beta61}, \ref{tab:beta60}, \ref{tab:beta73}, and
\ref{tab:beta82} show the complete set of possible p-shell states for 
$^6$He, $^6$Li, $^7$He, and $^8$He, respectively.
Tables~\ref{tab:beta71}, \ref{tab:beta81}, and \ref{tab:beta80} show the
lowest-lying p-shell states for $^7$Li, $^8$Li, and $^8$Be.
In previous work \cite{WPCP00}, only the two most spatially symmetric $LS[n]$
components were included in $^8$Li and $^8$Be, but now all are included.
The added components have little effect on the energies of the lowest states 
of given $(J^\pi,T)$, but their inclusion can be important for higher excited 
states; they are also expected to be important for some electroweak
transitions.  The diagonalizations were done for the AV18/UIX Hamiltonian
and the resulting $\beta_{LS[n]}$ used for all the Hamiltonians reported here;
a few tests have shown insignificant dependence of the $\beta_{LS[n]}$ 
on $V_{ijk}$.

Additional improvements in the trial wave have been made by refining the
detailed shapes of the $u_p(r)$, $f_{sp}(r)$, and $f_{pp}(r)$ functions; 
these improvements are described in more detail with the numerical results
in Sec.V.

\section {GFMC methods for ground states}

Green's function Monte Carlo calculations of light nuclei have
previously been performed for ground states of light 
nuclei~\cite{PPCW95,PPCPW97,WPCP00,PPWC01,PVW02} and for the lowest-energy
states with distinct quantum numbers.  In this section we briefly review
the GFMC method as applied to light nuclei; Refs.~\cite{PPCPW97,WPCP00,PVW02}
should be consulted for detailed descriptions of the method and
tests of its reliability.
We then report 
new algorithmic developments which significantly increase the
speed of these calculations.  These
improvements result in a much more efficient treatment of $N\!N\!N$
interactions, particularly the very small, but computationally
expensive, terms when three nucleons are well separated.
In the next section we discuss the extensions required to treat 
excited states of the same quantum numbers.

Starting with the trial function $\Psi_T$, GFMC provides a 
means of computing a wave function propagated in imaginary time:
\begin{equation}
\Psi (\tau) = \exp [ - (H - E_0) \tau ] \Psi_T \ =\ 
\exp [ - (H - E_0) \Delta \tau]^n \Psi_T,
\end{equation}
by stringing together a series of small time ( $\Delta \tau = \tau / N$)
evolution operators.
The energy, $E(\tau)$, is evaluated using a ``mixed'' expectation value
between $\Psi_T$ and $\Psi(\tau)$:
\begin{equation}
E(\tau) = \langle H(\tau) \rangle_{\rm Mixed} =
\frac{\langle \Psi(\tau) | H | \Psi_T\rangle}
     {\langle \Psi(\tau)   |   \Psi_T\rangle} \ .
\label{eq:mixed}
\end{equation}
In principle, $E(\tau)$ is an upper bound to the energy of the lowest 
eigenstate that is not orthogonal to $\Psi_T$.

The $\Psi (\tau)$ is represented by a vector function of ${\bf R}$,
where the components of the vector are the amplitudes of each
of the distinct spin-isospin states
in the system. The propagator
\begin{equation}
G_{\alpha,\beta} ({\bf R},{\bf R'}; \Delta \tau )
=   \langle {\bf R}, \alpha |
\exp [ - (H - E_0) \Delta \tau ] | {\bf R'}, \beta \rangle
\end{equation}
depends upon the 6A spatial coordinates ${\bf R}$ and ${\bf R'}$,
as well as the spin-isospin states $\alpha$ and $\beta$. 
It is calculated with leading errors of $(\Delta \tau)^3$.
The full wave function is then obtained as a product over many
small time steps:
\begin{eqnarray}
\Psi ({\bf R}_n, \tau)
&=& \int G ({\bf R}_n, {\bf R}_{n-1}) \cdots G ({\bf R}_1, {\bf R}_{0}) \nonumber \\
&\times& \Psi_T({\bf R}_{0}) d {\bf R}_{n-1} \cdots d{\bf R}_0,
\end{eqnarray}
where we have omitted the spin-isospin labels for simplicity.

A central ingredient in a GFMC calculation is the propagator 
$G({\bf R},{\bf R'}; \Delta \tau )$.  
It is desirable to use as large a time step $\Delta \tau$
as possible to
speed up the calculation, however the maximum time step is limited by the
accuracy with which we can evaluate the propagator 
$G ({\bf R},{\bf R'}; \Delta \tau )$.
The short-time propagator typically used for $N\!N$ interactions is
\begin{eqnarray}
&&G_{\alpha,\beta} ({\bf R},{\bf R'}; \Delta \tau ) = \nonumber  \\
&&G_0 ({\bf R},{\bf R'} ) 
 \langle \alpha | 
\left[ {\cal S} \prod_{i<j} \frac 
{g_{ij} ({\bf r}_{ij}, {\bf r'}_{ij})}
{g_{0,ij} ({\bf r}_{ij}, {\bf r'}_{ij})}
\right]
| \beta \rangle,
\end{eqnarray}
where the free-particle propagator $G_0$ and the free two-nucleon
propagator $g_{0,ij}$ are simple Gaussians~\cite{PPCW95}, and the interacting
pair propagator,
\begin{equation}
g_{ij} ( 
{\bf r}_{ij},
{\bf r'}_{ij}; \Delta \tau) =
\langle {\bf r}_{ij} | \exp [ - \Delta \tau (T_{ij} + v_{ij}) ]
| {\bf r}'_{ij} \rangle  ,
\end{equation}
is easily calculable from the two-body potential $v_{ij}$ and the relative
kinetic energy $T_{ij}$. We calculate the propagator with the simplified
AV8' interaction, the difference between the full Hamiltonian H and
the simplified H' is treated perturbatively. \cite{PPCPW97}

Because of the large number of spin-isospin amplitudes which must
be evaluated, the overall speed of the computation for larger nuclei
is dominated by the calculation of the $N\!N\!N$ propagator.
This term was originally treated in a 
very straightforward manner.  Including $N\!N\!N$ interactions, the
full propagator $G$ can be calculated as:
\begin{widetext}
\begin{eqnarray}
G_{\alpha,\beta} ({\bf R},{\bf R'}; \Delta \tau ) & = &  \exp [ E_0 \tau] \ 
G_0 ({\bf R},{\bf R'} ) \ 
\exp [ - \sum ( V_{ijk}^R ({\bf R})
+ V_{ijk}^R ({\bf R'}))\frac{\Delta \tau}{2}]  \nonumber \\
&\times& \langle \alpha | I_3 ({\bf R}; \frac{\Delta \tau}{2}) | \gamma \rangle
\langle \gamma | 
\left[ {\cal S} \prod_{i<j} \frac 
{g_{ij} ({\bf r}_{ij},{\bf r}'_{ij})}
{g_{0,ij}({\bf r}_{ij}, {\bf r}'_{ij})}
\right]
| \delta \rangle
\langle \delta | I_3 ({\bf R'}; \frac{\Delta \tau}{2}) | \beta \rangle,
\label{eq:fullprop}
\end{eqnarray}
\end{widetext}
where
\begin{equation}
I_3 ({\bf R}; \Delta \tau / 2 ) = 
\left[ 1 - \frac{\Delta \tau}{2} \sum V_{ijk}^{\pi} ( {\bf R})\right].
\label{eq:ithree}
\end{equation}
Here $V_{ijk}^R$ is the spin-isospin independent part of the 
$N\!N\!N$ interaction, which can be trivially exponentiated.
The $V_{ijk}^{\pi}$ contains the two-pion-exchange component of the
$N\!N\!N$ interaction, and in the case of the Illinois potentials,
the three-pion-exchange component.

We have previously employed the fact that the largest part of
$V_{ijk}^{\pi}$ is a sum of products of spin and isospin anticommutators that
can be expressed as
a sum of terms each of which contain only two-nucleon spin and isospin 
operators.  This part, which we denote $\tilde{V}_{ijk}$,
is not much more difficult to treat than the typical $N\!N$ potential,
i.e., while it involves spatial coordinates of three particles, the
spin-isospin algebra is equivalent to that of an $N\!N$ interaction.

The remaining terms
depend upon the spins and isospins
of three nucleons and are comparatively weak. 
If we define a propagator $\tilde G$ which is obtained
from the $N\!N$ potential $v_{ij}$ and the $\tilde{V}_{ijk}$
through Eq.~(\ref{eq:fullprop}), the complete propagator 
including the full $N\!N\!N$ interaction can be written as:
\begin{widetext}
\begin{eqnarray}
G_{\alpha,\beta} ({\bf R}_M,{\bf R}_1; M \Delta \tau ) & = &
 [ 1 - (V_{ijk} ({\bf R}_M) - \tilde{V}_{ijk} ({\bf R}_M)) M \frac{\Delta \tau}{2}]
\nonumber \\
&\times& \left[ \prod 
\tilde{G} ({\bf R}_M, {\bf R}_{M-1}; \Delta \tau )
...
\tilde{G} ({\bf R}_2, {\bf R}_{1}; \Delta \tau )
\right] 
 [ 1 - (V_{ijk} ({\bf R}_1) - \tilde{V}_{ijk} ({\bf R}_1)) M \frac{\Delta \tau}{2}],
\label{eq:tnifull}
\end{eqnarray}
\end{widetext}
where the number $M$ of intermediate steps with the 
simplified propagator $\tilde{G}$ is typically four to five.
This treatment is equivalent to evaluating the propagation
with a simplified Hamiltonian containing only two-body spin and
isospin operators, and then correcting the end points for the  difference
between the full $N\!N\!N$ interaction and the simplified one.
Clearly this method has the same short-time limit as the original
implementation, but the computational time is significantly decreased because
the number of full three-nucleon spin-isospin operations is much
reduced.

Further generalizing this method, it is possible
to introduce artificial fluctuations into the propagator
which average to the correct propagator in the short-time limit,
but are much more efficient computationally.
We can replace the factor $I_3 ({\bf R}) $ above  (Eq. \ref{eq:ithree})
by
\begin{eqnarray}
I_3 ({\bf R})  & = &  1 - \frac{\Delta \tau}{2}   \nonumber\\
&\times& \sum_{i<j<k} \int^1_0 dq_{ijk}
\ \theta ( P_{ijk} - q_{ijk})  \ \frac{V_{ijk}}{P_{ijk}} ,
\label{eq:I3}
\end{eqnarray}
where $\theta (x < 0) = 0$ and $\theta (x > 0 ) = 1$.
The function $P_{ijk}$ is an arbitrary function of the particle coordinates
$( {\bf r}_i, {\bf r}_j, {\bf r}_k )$ subject to the condition 
$ 0 \leq P_{ijk} \leq 1$.
Integrating over the ``auxiliary fields'' $q_{ijk}$ trivially recovers
the original term in the propagator, as with probability $P_{ijk}$
the triplet term is $V_{ijk} / P_{ijk}$ and with probability 
$1-P_{ijk}$ the triplet contribution is zero.  The original formulation
for the propagator, Eq.~(\ref{eq:ithree}), is equivalent to
choosing $P_{ijk} = 1$.

Choosing a $P_{ijk}$ which decreases as the three-nucleon separation
increases is computationally very advantageous, however.   Based upon
various trials, we have used
\begin{eqnarray}
 P_{ijk} & = & \left\{ \begin{array}{cl}
  1 ,                      &  X_{\mbox{max}} < X_{ijk} ,  \\
  X_{ijk}/X_{\mbox{max}} , &  X_{\mbox{min}} \leq  X_{ijk} \leq X_{\mbox{max}} , \\
  X_{\mbox{min}} ,         &  X_{ijk} < X_{\mbox{min}} ,
  \end{array}  \right. 
\end{eqnarray}
where
\begin{equation}
 X_{ijk} = \tilde{T}(r_{ij})\tilde{T}(r_{ik}) + \tilde{T}(r_{ij})\tilde{T}(r_{jk}) + 
   \tilde{T}(r_{ik})\tilde{T}(r_{jk}) ,
\end{equation}
\begin{equation}
  \tilde{T}(r)  =   \left\{ \begin{array}{cl}
   T(m_\pi r) ,              &  r > r_{\mbox{max}} ,  \\
   T(m_\pi r_{\mbox{max}}) , & r \leq  r_{\mbox{max}} . 
  \end{array}  \right. 
\end{equation}
Here $T(x)$ is the radial tensor function used in $V_{ijk}$ (see
Eqs.~(3.5-8) of Ref.~\cite{PPWC01}) and $r_{\mbox{max}}$ is the
location of the maximum of $T(m_\pi r)$.  
For small triplet separations, as effectively specified by $X_{\mbox{max}}$,
the full contribution is always
calculated.  For larger separations, the contribution is calculated
less often.  Because of the $1/P_{ijk}$ in Eq.~(\ref{eq:I3}), one does not
want $P_{ijk}$ to become too small; $X_{\mbox{min}}$ is used to control this.
Typical values are $X_{\mbox{max}}=5.0$ and $X_{\mbox{min}}=0.01~$.

This technique essentially invokes a larger time step
for well-separated triplets.   Tests are of course required 
to make sure that the overall time step is small enough to
avoid physically significant errors in the expectation values
of interest.  This method is particularly important in
nuclear physics applications, where the triplet computations are
so expensive, but could also be employed in other systems where
quantum Monte Carlo methods are used.

We have also adopted a similar scheme in computing the 
quadratic spin-orbit $(L \cdot S)^2$ and angular momentum ($L^2$) 
contributions to the energy in both
VMC and GFMC calculations.  These terms in the interaction are
typically quite small at large pair separations, 
and so we can evaluate them with a
probability $P_{ij}$ and then multiply the calculated contribution
with the inverse of $P_{ij}$. Using this scheme enables one to
calculate the energies and expectation values in the AV18 interaction
at a computational cost much closer to evaluating the 
simplified AV8$^\prime$ interaction.

\section{ GFMC evaluation of excited states}

It is possible to treat at least a few excited states with the
same quantum numbers using VMC and GFMC methods.
The VMC calculations have been described above, and essentially
involve solving a generalized eigenvalue problem, Eqs.~(\ref{eq:geneigen-e})
and (\ref{eq:geneigen-n}).
The same basic method can be applied in GFMC calculations, though
the implementation is slightly more involved.
In this section, $\Psi_{T,i}$ represents the trial wave function for
the $i^{th}$ state of specified $(J^\pi,T)$ and $\Psi_i(\tau)$ is
the GFMC wave function propagated from it.  By construction
$\langle \Psi_{T,i} | \Psi_{T,j} \rangle = 0$ for $i \neq j$.
We would like to calculate the Hamiltonian and
normalization matrix elements as a function of $\tau$:
\begin{eqnarray}
H_{ij}(\tau) & = & \frac {\langle \Psi_i(\tau/2)| \ H \ | \Psi_j(\tau/2) \rangle}
 {|\Psi_i(\tau/2)| |\Psi_j(\tau/2)|} ,  \\
N_{ij} (\tau) & = & \frac {\langle \Psi_i(\tau/2) | \Psi_j(\tau/2) \rangle}
 {|\Psi_i(\tau/2)| |\Psi_j(\tau/2)|} ,
\end{eqnarray}
where $| \Psi_i | = | \langle \Psi_i | \Psi_i \rangle |^{1/2}$.
Solving the generalized eigenvalue problem with
these Hamiltonian and normalization matrix elements would yield
improved upper bounds for the ground and low-lying excited states
of the system.  In the limit $\tau \rightarrow \infty $ the
solutions would be exact.

In GFMC we can compute mixed expectation values such as
\begin{equation}
 \tilde{O}_{ij}(\tau) = \frac {\langle \Psi_i(\tau)| \ O \ | \Psi_{T,j} \rangle}
 {\langle \Psi_i(\tau)| \Psi_{T,i} \rangle} ,
\end{equation}
where the denominator involves just state $i$.  
Since the propagator commutes with the Hamiltonian, the desired matrix
elements can be computed as:
\begin{eqnarray}
H_{ij} (\tau) & = & [ \tilde{H}_{ij} \tilde{H}_{ji} ]^{\case{1}{2}} , \\
N_{ij} (\tau) & = & [ \tilde{N}_{ij} \tilde{N}_{ji} ]^{\case{1}{2}} ,
\end{eqnarray}
where we use expectation values computed from separately propagated $\Psi_i(\tau)$
and $\Psi_j(\tau)$.  For $i=j$ these equations reduce to the standard
GFMC calculation described in Sec.~III.

For larger nuclei, $A \geq 8$, the fermion sign problem, i.e., the problem
of symmetric noise, becomes a sufficiently large computational burden, that a
``constrained path'' algorithm must be utilized~\cite{WPCP00}.
In this case the
propagator for different states is not identical, as it involves constraints
based upon the different trial wave functions, and the above equations
are only approximate.
As is described in \cite{WPCP00} for calculations of the lowest state of
given ($J^\pi$, T), any errors in the energy that are introduced
by the constrained path can be removed by doing 10 to 20 unconstrained steps
before evaluating the energy.  
In the present work we follow this same proceedure and make
10 to 20 unconstrained steps before evaluating $H_{ij}$ and $N_{ij}$.

\begin{figure}[ht!]
\centering
\includegraphics[width=2.5in,angle=270]{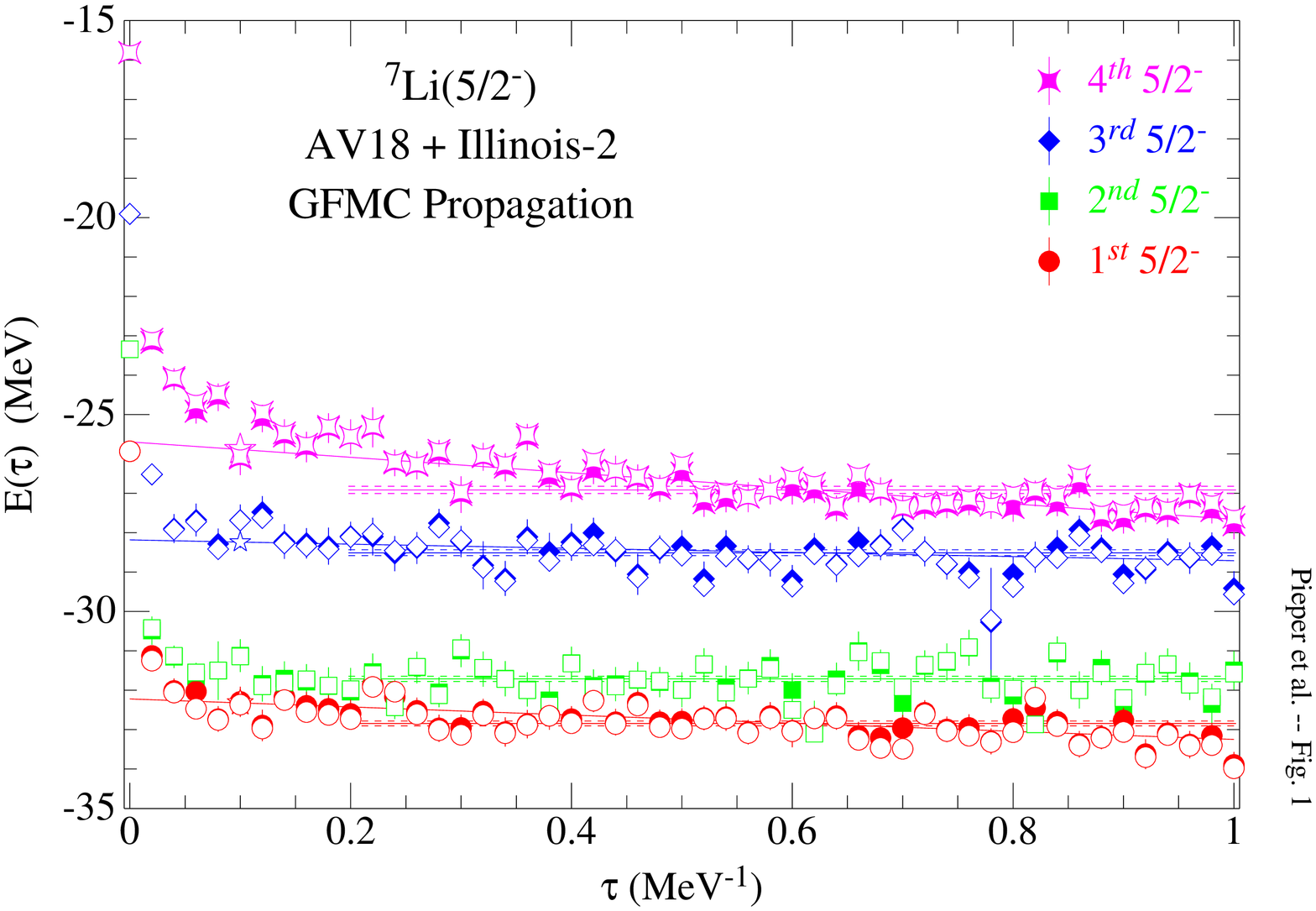}
\caption{(Color online)
GFMC energies of four $\frac{5}{2}^-$ states in $^7$Li versus imaginary time, 
$\tau$.  The solid symbols show the computed energies at each $\tau$; open 
symbols show the results of the rediagonalization discussed in the text.}
\label{fig:e-vs-tau}
\end{figure}
\begin{figure}[ht!]
\centering
\includegraphics[width=2.3in,angle=270]{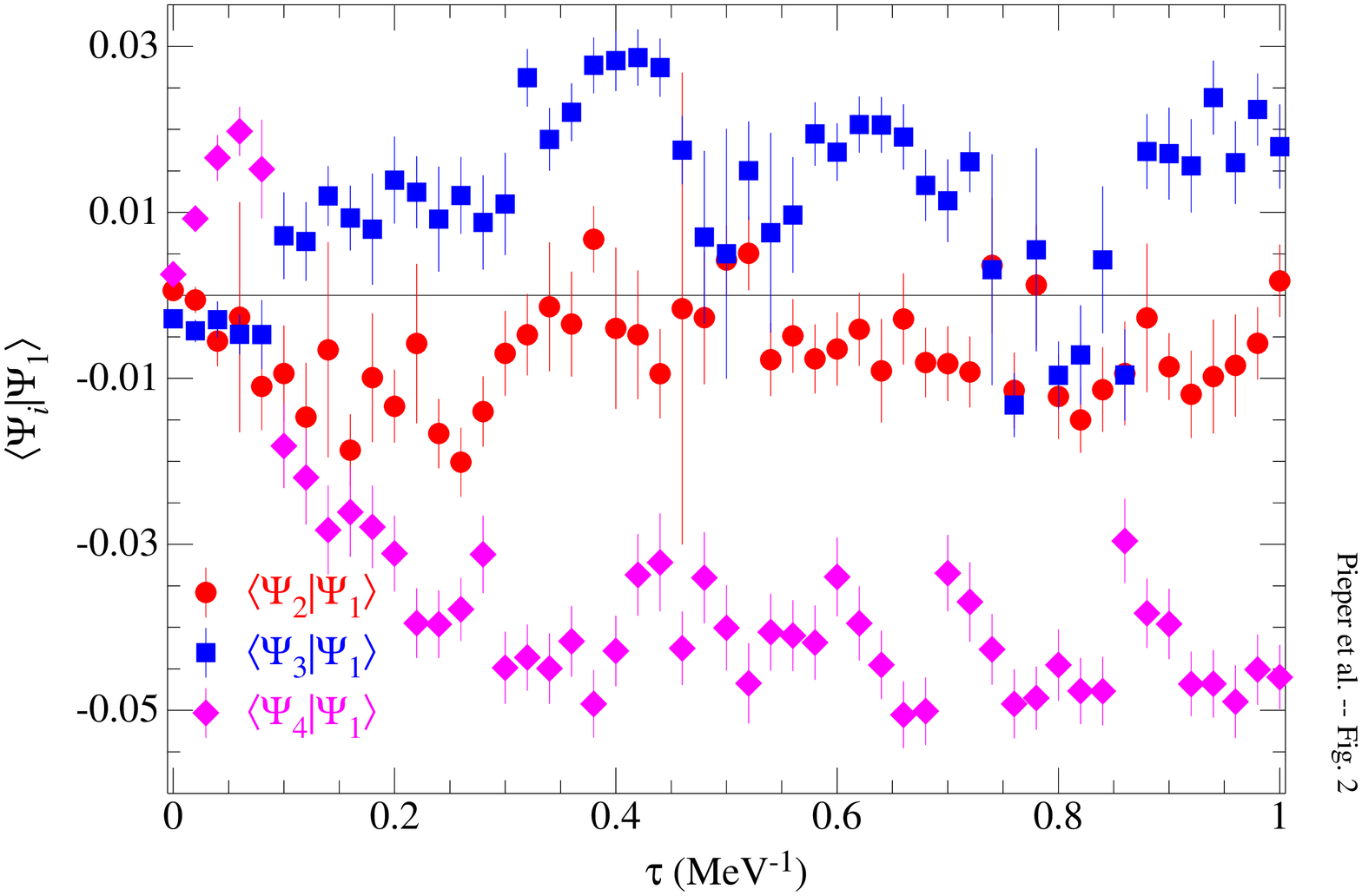}
\caption{(Color online)
Overlaps of the GFMC wave functions for the first $\frac{5}{2}^-$ state in 
$^7$Li with the other three $\frac{5}{2}^-$ states versus imaginary time.}
\label{fig:overlap-vs-tau}
\end{figure}

As an example, Figs. \ref{fig:e-vs-tau} and \ref{fig:overlap-vs-tau} show 
the computation of the energies of four $\case{5}{2}^-$ states in $^7$Li;
the $\Psi_{T,i}$ used to start the propagations are defined in Table~\ref{tab:beta71}.
The diagonal $H_{ii}(\tau)$ are shown as solid symbols in Fig.~\ref{fig:e-vs-tau}.
The lowest state has mainly [43] symmetry and can easily decay to the
$\alpha$+t channel; it has a large experimental width (918 keV) and its
computed energy is slowly decreasing to the energy of the separated clusters.
The remaining states are mainly of [421] symmetry and so are principally
connected to the $^6$Li+n channel.  The second $\case{5}{2}^-$ state is
experimentally just above the $^6$Li+n threshold and has a small
width (80 keV); its computed energy becomes constant with increasing $\tau$.
The last two $\case{5}{2}^-$ states are not experimentally known,
but the very slow decrease with $\tau$ of the energy of the third state
suggests that this state may also be narrow.  Figure~\ref{fig:overlap-vs-tau}
shows the off-diagonal overlaps $N_{i1}(\tau)$; they are
small and do not show signs of steadily increasing with increasing $\tau$.
The $N_{i\neq j}(\tau\!=\!0)$ are not identically zero because the 
diagonalization that determined the $\beta_{LS[n]}$ was made in a
different Monte Carlo walk from the ones generating the $N_{ij}$.
These results show that the (constrained) GFMC propagation largely
retains the orthogonality of the starting $\Psi_{T,i}$.
Contrary to what might have been expected, the propagation of the
higher states does not quickly collapse to the lowest state.

The open symbols in Fig.~\ref{fig:e-vs-tau} show the results of
generalized eigenvalue calculations made for each $\tau$ using the
$H_{ij}(\tau)$ and $N_{ij}(\tau)$.  As is expected from the small
overlaps, there is no statistically significant change of the
rediagonalized energies from the directly computed $H_{ii}(\tau)$.
Based on this result and the generally constant $H_{ii}(\tau)$ that we
obtain for other cases,  we use the $H_{ii}(\tau)$ directly without
making a final rediagonalization.

\section {Numerical results for $A$ = 6 to 8 nuclei}

Figure \ref{fig:e-vs-tau} shows that for some unbound cases the GFMC
energy never stabilizes, but rather slowly decays with increasing
$\tau$.  In these cases a linear fit with non-zero slope provides a much
lower $\chi^2$ fit to the $E(\tau)$ for large $\tau$ than does a
constant fit.  We use this linear fit to extrapolate back to $\tau
\approx 0.08$ because it appears that the GFMC calculations for bound
systems have all become stable by $\tau \approx 0.08$.  Such
extrapolated values are printed in italics in the tables; they must be
considered as less reliable than the non-italicized values.

\begin{table}[ht!]
\caption{Energies and excitation energies (MeV) of selected nuclear
states calculated for the AV18/UIX Hamiltonian by VMC and GFMC.
Monte Carlo statistical errors in the last digits are shown in parentheses.
See the text for italicized GFMC values.}
\begin{tabular}{lcccc}
     & \multicolumn{2}{c}{Energy} & \multicolumn{2}{c}{Excitation Energy} \\
                                  &  VMC        &    GFMC        &    VMC  &    GFMC      \\
\hline                                                                         
$^4$He(0$^+$)                     & $-$27.39(3) &    $-$28.31(2) &         &              \\
\vspace*{-.10in}\\
$^6$He(0$^+$)                     & $-$24.79(9) &    $-$28.02(9) &         &              \\
$^6$He(2$^+$)                     & $-$23.07(9) &    $-$26.08(9) & ~1.7(1) &      ~1.9(1) \\
$^6$He(1$^+$)                     & $-$20.72(9) &\em $-$23.9(2)~ & ~4.1(1) &\em   ~4.2(2) \\
\vspace*{-.10in}\\
$^6$Li(1$^+$)                     & $-$27.96(8) &    $-$31.25(8) &         &              \\
$^6$Li(3$^+$)                     & $-$25.05(9) &    $-$28.45(8) & ~2.9(1) &      ~2.8(1) \\
$^6$Li(2$^+$)                     & $-$23.8(1)~ &    $-$27.25(8) & ~4.2(1) &      ~4.0(1) \\
$^6$Li(1$^+$, 2$^{nd}$)           & $-$22.69(8) &\em $-$26.2(1)~ & ~5.3(1) &\em   ~5.1(1) \\
\vspace*{-.10in}\\
$^7$He($\case{3}{2}^-$)           & $-$20.93(8) &    $-$26.3(1)~ &         &              \\
$^7$He($\case{1}{2}^-$)           & $-$18.8(1)~ &    $-$25.2(1)~ & ~2.1(1) &      ~1.1(1) \\
$^7$He($\case{5}{2}^-$)           & $-$18.33(9) &\em $-$23.9(1)~ & ~2.6(1) &\em   ~2.5(2) \\
\vspace*{-.10in}\\
$^7$Li($\case{3}{2}^-$)           & $-$33.0(3)~ &    $-$37.5(1)~ &         &              \\
$^7$Li($\case{1}{2}^-$)           & $-$32.9(1)~ &    $-$37.6(1)~ & ~~.1(3) &    ~$-$.1(2) \\
$^7$Li($\case{7}{2}^-$)           & $-$27.2(1)~ &    $-$32.2(1)~ & ~5.8(3) &      ~5.3(1) \\
$^7$Li($\case{5}{2}^-$)           & $-$26.61(9) &\em $-$31.1(1)~ & ~6.4(3) & \em  ~6.5(2) \\
$^7$Li($\case{5}{2}^-$, 2$^{nd}$) & $-$23.7(1)~ &    $-$29.7(2)~ & ~9.3(3) &      ~7.8(2) \\
$^7$Li($\case{3}{2}^-$, 2$^{nd}$) & $-$22.8(1)~ &    $-$29.1(2)~ & 10.2(3) &      ~8.5(3) \\
$^7$Li($\case{7}{2}^-$, 2$^{nd}$) & $-$21.8(1)~ &    $-$28.1(2)~ & 11.2(3) &      ~9.5(2) \\
$^7$Li($\case{5}{2}^-$, 3$^{rd}$) & $-$20.5(1)~ &    $-$27.0(2)~ & 12.5(3) &      10.5(2) \\
$^7$Li($\case{5}{2}^-$, 4$^{th}$) & $-$16.8(2)~ &\em $-$24.4(5)~ & 16.2(4) &\em   13.1(5) \\
\vspace*{-.10in}\\
$^8$He(0$^+$)                     & $-$20.75(7) &    $-$27.7(1)~ &         &              \\
$^8$He(2$^+$)                     & $-$18.2(1)~ &    $-$25.0(1)~ & ~2.5(1) &      ~2.8(2) \\
$^8$He(1$^+$)                     & $-$17.56(8) &\em $-$23.3(3)~ & ~3.2(1) &\em   ~4.4(3) \\
\vspace*{-.10in}\\
$^8$Li(2$^+$)                     & $-$30.7(1)~ &    $-$38.8(1)~ &         &              \\
$^8$Li(1$^+$)                     & $-$29.6(1)~ &    $-$37.8(2)~ & ~1.1(1) &      ~1.0(2) \\
$^8$Li(0$^+$)                     & $-$28.6(1)~ &    $-$36.9(1)~ & ~2.1(1) &      ~2.0(2) \\
$^8$Li(3$^+$)                     & $-$27.24(9) &    $-$35.4(1)~ & ~3.5(1) &      ~3.4(2) \\
$^8$Li(1$^+$, 2$^{nd}$)           & $-$27.7(1)~ &    $-$35.2(1)~ & ~3.0(2) &      ~3.6(2) \\
$^8$Li(4$^+$)                     & $-$24.42(9) &    $-$32.3(2)~ & ~6.3(1) &      ~6.5(2) \\
\vspace*{-.10in}\\
$^8$Be(0$^+$)                     & $-$48.6(1)~ &    $-$55.2(1)~ &         &              \\
$^8$Be(2$^+$)                     & $-$45.71(9) &    $-$52.1(2)~ & ~2.9(1) &      ~3.1(3) \\
$^8$Be(4$^+$)                     & $-$37.59(9) &\em $-$44.2(5)~ & 11.0(1) &\em   11.0(5) \\
$^8$Be(1$^+$)                     & $-$28.07(8) &    $-$37.0(2)~ & 20.6(1) &      18.2(2) \\
$^8$Be(2$^+$, 2$^{nd}$)           & $-$28.8(1)~ &\em $-$36.4(6)~ & 19.8(1) &\em   18.9(6) \\
$^8$Be(3$^+$)                     & $-$26.69(7) &    $-$35.2(2)~ & 21.9(1) &      20.0(2) \\
\end{tabular}
\label{tab:uix-vg}
\end{table}

Selected VMC and GFMC energies computed for the
AV18/UIX Hamiltonian are shown in Table~\ref{tab:uix-vg}.
The VMC energies for the $\Psi_T$ of Eq.~(\ref{eq:psit}) are significantly
improved compared to those reported in Refs.~\cite{PPCPW97,WPCP00}.
The $^4$He energy has been lowered by 0.4 MeV, the $A$=6 energies by 1 MeV,
the $A$=7 energies by 2 MeV, and the $A=8$ energies by 2.75 MeV.
Compared to the more sophisticated (and more expensive) $\Psi_V$ 
of Refs.~\cite{PPCPW97,WPCP00}, which 
includes spin-orbit and additional three-body correlations, the energies 
from the present $\Psi_T$ are as good for $A$=6 and even better for $A$=7,8.
\begin{table*}[ht!]
\caption{$A$ = 4, 6 and 7 GFMC energies for the AV8$^\prime$, AV18, and 
AV18/IL2 Hamiltonians
compared with experimental values~\cite{exp567}. 
Except as noted, states have the same isospin as the corresponding ground state.
Square brackets enclose experimental energies which have uncertain $J^\pi$ assignments.
See the text for italicized GFMC values.  All values are in MeV.}
 
\begin{tabular}{lccccc}
                                  &   AV8$^\prime$ &        AV18    &     AV18/IL2   & \multicolumn{2}{c}{Experiment} \\
                                  &                &                &                &  Energy      & Width \\
\hline                                                                  
$^4$He(0$^+$)                     &    $-$25.14(2) &    $-$24.07(4) &    $-$28.37(3) & $-$28.30(0)~ &       \\
\vspace*{-.10in}\\
$^6$He(0$^+$)                     &    $-$25.11(3) &    $-$23.8(1)~ &    $-$29.28(2) & $-$29.27~~~~ &       \\
$^6$He(2$^+$)                     &    $-$23.21(4) &    $-$21.9(1)~ &    $-$27.3(2)~ & $-$27.47(3)~ & 0.113 \\
$^6$He(2$^+$, 2$^{nd}$)           &\em $-$21.4(1)~ &\em $-$20.4(1)~ &\em $-$24.6(1)~ &              &       \\
$^6$He(1$^+$)                     &\em $-$21.0(1)~ &\em $-$19.6(2)~ &    $-$24.5(1)~ &              &       \\
$^6$He(0$^+$, 2$^{nd}$)           &    $-$20.15(6) &    $-$19.0(1)~ &    $-$23.3(1)~ &              &       \\
\vspace*{-.10in}\\                                                                      
$^6$Li(1$^+$)                     &    $-$28.15(3) &    $-$26.9(1)~ &    $-$32.0(1)~ & $-$31.99~~~~ &       \\
$^6$Li(3$^+$)                     &    $-$25.33(3) &    $-$24.1(1)~ &    $-$29.8(1) & $-$29.80~~~~ & 0.024 \\
$^6$Li(0$^+$, $T$=1)              &    $-$24.31(3) &    $-$23.0(1)~ &    $-$28.6(1)~ & $-$28.43~~~~ &       \\
$^6$Li(2$^+$)                     &    $-$24.18(5) &    $-$22.8(1)~ &    $-$27.8(1)~ & $-$27.68(2)~ & 1.3~~ \\
$^6$Li(2$^+$, $T$=1)              &    $-$22.48(4) &    $-$21.1(1)~ &    $-$26.5(1)~ & $-$26.62~~~~ & 0.54~ \\
$^6$Li(1$^+$, 2$^{nd}$)           &    $-$23.19(6) &    $-$22.0(1)~ &\em $-$26.4(2)~ & $-$26.34(5)~ & 1.5~~ \\
$^6$Li(1$^+$, 3$^{rd}$)           &\em $-$19.7(2)~ &    $-$18.8(1)~ &    $-$23.4(1)~ &              &       \\
\vspace*{-.10in}\\                                                                      
$^7$He($\case{3}{2}^-$)           &\em $-$23.39(7) &\em $-$21.9(1)~ &\em $-$28.8(2)~ & $-$28.82(3)~ & 0.16~ \\
$^7$He($\case{1}{2}^-$)           &\em $-$22.05(9) &\em $-$20.7(1)~ &\em $-$25.9(2)~ &              &       \\
$^7$He($\case{5}{2}^-$)           &\em $-$20.8(1)~ &\em $-$19.4(1)~ &    $-$25.5(1)~ &[$-$25.90(10)]& 1.99  \\
$^7$He($\case{3}{2}^-$, 2$^{nd}$) &                &                &    $-$25.0(1)~ &              &       \\
$^7$He($\case{3}{2}^-$, 3$^{rd}$) &                &                &\em $-$21.7(3)~ &              &       \\
\vspace*{-.10in}\\                                                                      
$^7$Li($\case{3}{2}^-$)           &    $-$33.86(4) &    $-$32.0(1)~ &    $-$38.9(1) & $-$39.24~~~~ &       \\
$^7$Li($\case{1}{2}^-$)           &    $-$33.95(4) &    $-$32.2(1)~ &    $-$38.7(1)~ & $-$38.76~~~~ &       \\
$^7$Li($\case{7}{2}^-$)           &\em $-$28.6(1)~ &\em $-$26.8(2)~ &    $-$34.0(1)~ & $-$34.59(1)~ & 0.069 \\
$^7$Li($\case{5}{2}^-$)           &    $-$28.11(6) &    $-$26.4(1)~ &\em $-$32.3(1)~ & $-$32.64(5)~ & 0.92~ \\
$^7$Li($\case{5}{2}^-$, 2$^{nd}$) &    $-$26.42(7) &    $-$24.5(1)~ &    $-$31.7(1)~ & $-$31.79~~~~ & 0.08~ \\
$^7$Li($\case{3}{2}^-$, 2$^{nd}$) &\em $-$25.4(2)~ &    $-$24.4(1)~ &    $-$29.7(1)~ & $-$30.49~~~~ & 4.7~~ \\
$^7$Li($\case{1}{2}^-$, 2$^{nd}$) &    $-$25.37(7) &    $-$23.8(1)~ &    $-$29.1(2)~ & $-$30.15~~~~ & 2.8~~ \\
$^7$Li($\case{7}{2}^-$, 2$^{nd}$) &    $-$24.59(7) &    $-$23.0(1)~ &    $-$29.2(2)~ & $-$29.67(10) & 0.44~ \\
$^7$Li($\case{5}{2}^-$, 3$^{rd}$) &                &                &    $-$28.6(1)~ &              &       \\
$^7$Li($\case{5}{2}^-$, 4$^{th}$) &                &                &\em $-$25.8(1)~ &              &       \\
\end{tabular}
\label{table:energies-67}
\end{table*}
\begin{table*}[ht!]
\caption{$A$ = 8 GFMC energies for the AV8$^\prime$, AV18, and AV18/IL2
Hamiltonians compared with experimental values~\cite{exp8}.
Conventions are the same as in Table~\protect\ref{table:energies-67}.
The $2^+$ states near $-39.7$~MeV in $^8$Be are strongly isospin mixed; their experimental
energies, marked with an asterisk, have not been corrected for this effect.}
 
\begin{tabular}{lccccc}
                                  &   AV8$^\prime$ &        AV18   &     AV18/IL2   & \multicolumn{2}{c}{Experiment} \\
                                  &                &               &                &  Energy      & Width \\
\hline                                                                 
$^8$He(0$^+$)                     &\em $-$24.3(1)~ &    $-$23.0(1) &    $-$31.72(4) & $-$31.41(1)~ &       \\
$^8$He(2$^+$)                     &    $-$21.93(6) &    $-$20.4(1) &    $-$27.0(2)~ & $-$28.31(50) & 0.6~~ \\
$^8$He(1$^+$)                     &\em $-$20.3(2)~ &\em $-$18.8(3) &    $-$25.9(2)~ &              &       \\
$^8$He(0$^+$, 2$^{nd}$)           &                &               &    $-$24.8(2)~ &              &       \\
$^8$He(2$^+$, 2$^{nd}$)           &                &               &    $-$23.7(2)~ &              &       \\
\vspace*{-.10in}\\                                                                     
$^8$Li(2$^+$)                     &    $-$34.74(6) &    $-$32.7(1) &    $-$41.9(2)~ & $-$41.28~~~~ &       \\
$^8$Li(1$^+$)                     &    $-$34.34(7) &    $-$32.1(1) &    $-$40.5(2)~ & $-$40.30~~~~ &       \\
$^8$Li(3$^+$)                     &    $-$32.16(7) &    $-$30.1(2) &    $-$39.4(2)~ & $-$39.02~~~~ & 0.033 \\
$^8$Li(0$^+$)                     &    $-$33.51(6) &    $-$31.5(1) &    $-$38.3(2)~ &              &       \\
$^8$Li(1$^+$, 2$^{nd}$)           &    $-$32.76(7) &    $-$31.0(1) &    $-$37.2(2)~ & $-$38.07~~~~ & 1.~~~ \\
$^8$Li(2$^+$, 2$^{nd}$)           &\em $-$31.2(2)~ &    $-$29.7(1) &    $-$36.6(4)~ &              &       \\
$^8$Li(2$^+$, 3$^{rd}$)           &    $-$30.49(7) &    $-$28.7(2) &    $-$36.9(2)~ &              &       \\
$^8$Li(1$^+$, 3$^{rd}$)           &    $-$31.14(7) &    $-$29.1(2) &    $-$35.9(2)~ &[$-$35.88]~~~ & 0.65~ \\
$^8$Li(3$^+$, 2$^{nd}$)           &    $-$28.47(7) &    $-$26.3(1) &    $-$33.8(2)~ &[$-$35.18(10)]& 1.~~~ \\
$^8$Li(4$^+$)                     &    $-$29.1(1)~ &    $-$27.1(1) &    $-$34.7(2)~ & $-$34.75(2)~ & 0.035 \\
\vspace*{-.10in}\\                                                                     
$^8$Be(0$^+$)                     &    $-$48.95(7) &    $-$46.3(2) &    $-$56.3(1)~ & $-$56.50~~~~ &       \\
$^8$Be(2$^+$)                     &\em $-$45.6(3)~ &    $-$43.7(2) &\em $-$52.2(2)~ & $-$53.44(3)~ & 1.4~~ \\
$^8$Be(4$^+$)                     &\em $-$37.7(5)~ &    $-$36.2(2) &    $-$45.4(3)~ & $-$45.15(15) & 3.5~~ \\
$^8$Be(2$^+$, $T$=1)              &    $-$33.40(6) &    $-$31.2(1) &    $-$40.4(2)~ & $-39.87^*$~~ & 0.108 \\
$^8$Be(2$^+$, 2$^{nd}$)           &    $-$33.41(9) &    $-$31.0(2) &    $-$39.8(3)~ & $-39.58^*$~~~& 0.074 \\
$^8$Be(1$^+$, $T$=1)              &    $-$33.03(7) &    $-$30.7(1) &    $-$39.0(2)~ & $-$38.86~~~~ & 0.011 \\
$^8$Be(1$^+$)                     &    $-$33.25(9) &    $-$30.8(2) &    $-$39.4(3)~ & $-$38.35~~~~ & 0.138 \\
$^8$Be(3$^+$, $T$=1)              &    $-$30.87(7) &    $-$28.8(2) &    $-$38.0(2)~ & $-$37.43~~~~ & 0.27~ \\
$^8$Be(1$^+$, 2$^{nd}$)           &                &               &    $-$37.3(3)~ &              &       \\
$^8$Be(3$^+$)                     &    $-$31.68(9) &    $-$29.7(2) &    $-$37.2(1)~ & $-$37.26(1)~ & 0.23~ \\
$^8$Be(4$^+$, 2$^{nd}$)           &                &               &    $-$37.6(3)~ & $-$36.64(5)~ & 0.7~~ \\
$^8$Be(2$^+$, 3$^{rd}$)           &    $-$31.49(9) &    $-$29.1(2) &    $-$37.0(3)~ & $-$36.40~~~~ & 0.88~ \\
$^8$Be(0$^+$, 2$^{nd}$)           &                &               &    $-$36.0(3)~ & $-$36.30~~~~ & 0.72~ \\
$^8$Be(3$^+$, 2$^{nd}$)           &                &               &    $-$34.0(3)~ & $-$35.00~~~~ & 1.~~~ \\
$^8$Be(2$^+$, 4$^{th}$)           &                &               &    $-$34.0(3)~ & $-$34.30~~~~ & 0.8~~ \\
\end{tabular}
\label{table:energies-8}
\end{table*}
\begin{figure*}[ht!]
\centering
\includegraphics[width=4.9in,angle=270]{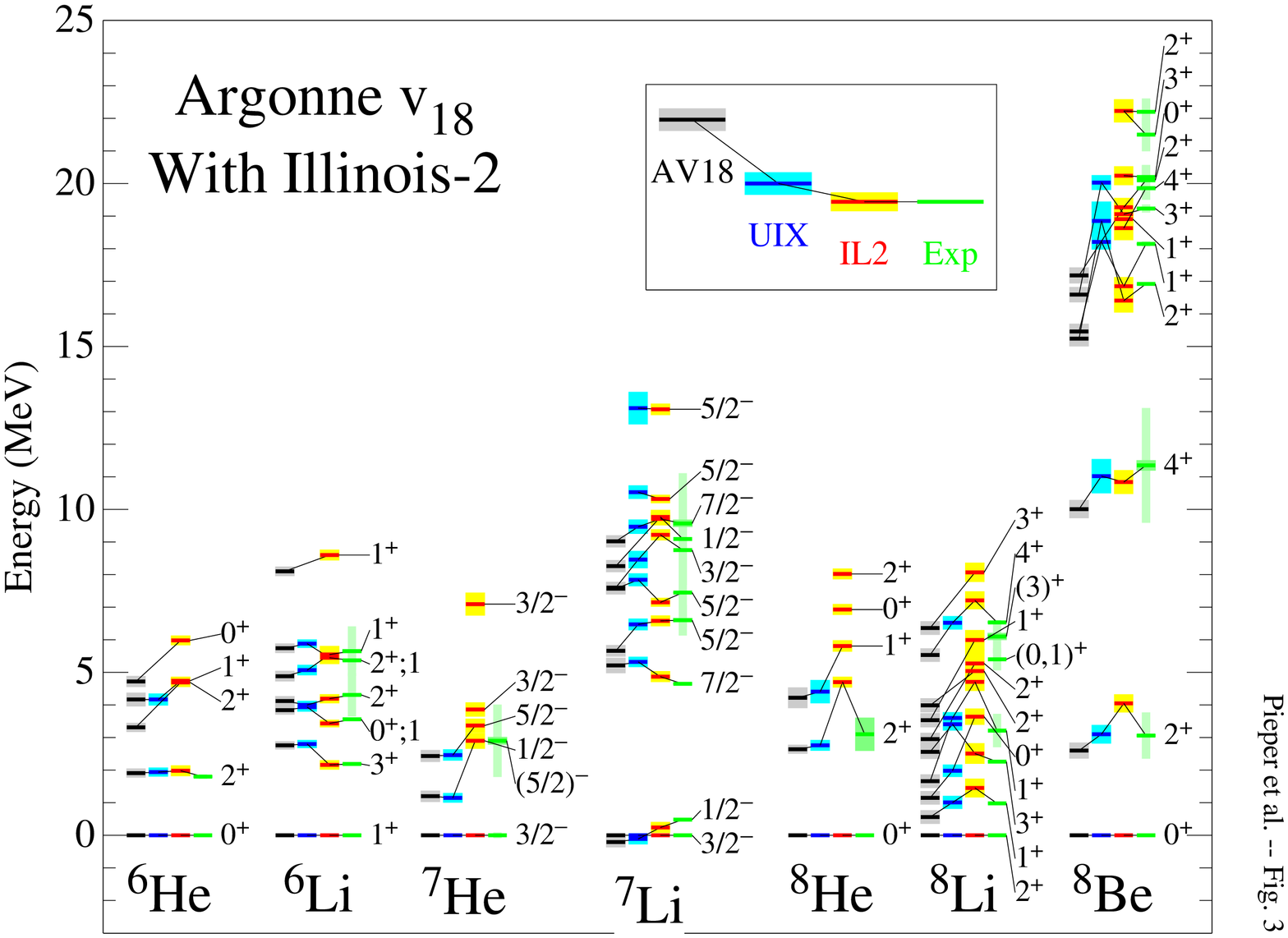}
\caption{(Color online)
GFMC excitation energies computed for the AV18, AV18/UIX, and AV18/IL2
Hamiltonians compared with experimental values.  The shaded bands show the 
Monte Carlo statistical (or experimental) errors.  The narrow shaded bars on
the experimental values show the experimental widths.}
\label{fig:estar}
\end{figure*}
The improvements are due primarily to 1) improved shapes for the two-body
tensor correlation functions $u_t(r)$, $u_{t\tau}(r)$ contained in $U_{ij}$
[Eq.~(\ref{eq:uij})], including allowing them to be different for s-shell and 
p-shell nuclei, 2) letting the $f_{sp}(r)$ vary with the $LS[n]$ wave
function component, and 3) using more extended $f_{sp}(r)$ and $f_{pp}(r)$ 
which allow the overall wave function to be more diffuse.
In particular, the lack of experimental charge radii for $A$=8 nuclei
makes it difficult to fix the optimal size of these wave functions; in 
the present case the matter radii for the ground states have been adjusted
to match the earlier GFMC results of Ref.~\cite{WPCP00}.  

Many of the GFMC results in Table~\ref{tab:uix-vg} have changed by 1 to
3 standard deviations from the results reported in Ref.~\cite{PPWC01}.
However, the quoted errors are statistical only; we have previously
estimated~\cite{PPCPW97,WPCP00} the systematic errors to be of the 
order of 1-2\%, and the changes reported here are within that range.
The changes are due to the improved treatment of $V_{ijk}$ discussed
above, propagation to larger $\tau$, and in some cases improved
$\beta_{LS[n]}$ choices.
These changes are significantly less than the corresponding
improvements in the $\Psi_T$ values,
indicating that (aside from modifications of the $\beta_{LS[n]}$)
high-energy contaminations have been removed from
the $\Psi_T$; these were previously easily corrected by the GFMC.

For the p-shell nuclei, the improved $\Psi_T$ energies are still
10 to 20\% higher than the GFMC values.  However the excitation energies
computed with the $\Psi_T$ are generally quite accurate, providing the
states are dominated by the same spatial symmetry as the ground state.  The
second $\case{5}{2}^-$ and higher states of $^7$Li and the
$1^+$ and $3^+$ states of $^8$Be have lower spatial symmetry than the 
corresponding
ground states and their VMC excitation energies are significantly too high; we
also see this pattern in the VMC energies of other higher states and
with other Hamiltonians.

Tables \ref{table:energies-67} and  \ref{table:energies-8} give GFMC
energies for the AV8$^\prime$, AV18, and AV18/IL2 Hamiltonians and 
also show the experimental energies and widths of the states that
have been observed (widths less than 0.01 MeV are not shown).  
Figure~\ref{fig:estar} shows the corresponding excitation energies.
The values
for second or higher states of a given $J^\pi$ are presented here
for the first time, however the other values have all been recomputed
with, in many cases, increased maximum $\tau$, improved determinations
of the $\beta_{LS[n]}$ in the $\Psi_T$, and
improved treatment of the $V_{ijk}$ in the propagation.  For these
reasons, some of the values are significantly different from 
previously published values.

The AV18/IL2 Hamiltonian was originally determined by making a
three-parameter fit to 17 states in the $A$ = 3 to 8
region~\cite{PPWC01}.  
There are 36 states with  $6\leq A\leq 8$ that have experimental energies in 
Tables \ref{table:energies-67} and \ref{table:energies-8}.
The RMS error in predicting the energies of these states
is only 0.60 MeV; the RMS error for just the
17 states with widths less than 0.2 MeV is 0.38 MeV and the RMS error
for the 7 ground-state energies is 0.31 MeV.  The RMS errors in
excitation energies are 0.76 MeV for all 29 excited states and 0.54 MeV
for the 10 narrow states.  
Analog states have been omitted from all of these averages but the states
with uncertain experimental $J^\pi$ assignments were included; omitting
the latter does not substantially change the above numbers.
These results are basically the same as the
RMS errors reported for 26 states in Ref.~\cite{PPWC01}.

\begin{table}[ht!]
\caption{GFMC expectation values of $V_{ijk}$ for the AV18/IL2 Hamiltonian}
 
\begin{tabular}{lcccc}
       & $V^{2\pi}$ & $V^{3\pi}$ & $V^R$   & $V_{ijk}$   \\
\hline
$^4$He & $-$16.3(1) &  ~~0.63(1) & ~7.3(1) & ~$-$8.4(1) \\
$^6$He & $-$18.9(1) & $-$0.58(1) & ~8.9(1) & $-$10.6(1) \\
$^6$Li & $-$20.1(2) & $-$0.09(3) & ~9.4(2) & $-$10.8(2) \\
$^7$He & $-$23.5(3) & $-$2.22(6) & 11.6(2) & $-$14.2(2) \\
$^7$Li & $-$24.9(2) & $-$0.50(3) & 11.7(1) & $-$13.7(2) \\
$^8$He & $-$27.0(1) & $-$4.36(2) & 13.8(1) & $-$17.5(1) \\
$^8$Li & $-$31.4(3) & $-$2.6(1)~ & 15.8(2) & $-$18.2(3) \\
$^8$Be & $-$35.9(2) &  ~~0.32(3) & 16.4(2) & $-$19.2(2) \\
\end{tabular}
\label{table:vijk}
\end{table}

Extensive breakdowns of the total energies were presented in
Ref.~\cite{PPWC01}.  The improved treatment of $V_{ijk}$ in the GFMC
propagation has resulted in significant changes to the expectation
values of $V_{ijk}$ and its components for the AV18/IL2 Hamiltonian;
revised values are shown in Table~\ref{table:vijk}.  The other
contributions to the total energies shown in Ref.~\cite{PPWC01}
are not significantly changed.  The isovector and isotensor
energies presented there have also not significantly changed in
the current calculations.  
Tables \ref{table:energies-67} and  \ref{table:energies-8} contain
energies of a few $T$=1 states for $^6$Li and $^8$Be; these have been
computed perturbatively using the isovector and isotensor energies
computed for $^6$He and $^8$Li, respectively.
Recently we have realized that there
is a significant difficulty in extracting precise RMS radii from
GFMC calculations, especially for weakly bound systems.  For this
reason we are deferring presenting updated values for the RMS radii.

\section {Conclusions}

We have demonstrated that it is possible to use GFMC to compute the
energies of multiple nuclear states with the same quantum numbers.  This
substantially increases the number of nuclear level energies that can be
compared to experimental values in the light p-shell region.  The AV18/IL2 
Hamiltonian presented in Ref.~\cite{PPWC01} gives a good description of 
this increased set of energies.  We are currently extending these 
calculations to higher excited states in the $A$=9,10 nuclei.  

We have also improved our treatment of the three-nucleon force in the
GFMC calculations.  This has altered the detailed breakdown of $N\!N\!N$
contributions as shown in Table~\ref{table:vijk}.  We expect a similar
alteration of these terms in the $A$=9,10 nuclei, but the total energies
should stay within 1-2\% of our previously reported values~\cite{PVW02}.

\acknowledgments

We thank D. Kurath, V. R. Pandharipande, and K.~Varga for useful discussions.
The many-body calculations were performed on the parallel computers of the
Laboratory Computing Resource Center, Argonne National Laboratory 
and the National Energy Research Scientific Computing Center.
The work of SCP and RBW is supported by the U. S. Department of Energy, 
Office of Nuclear Physics, under contract No. W-31-109-ENG-38.
The work of JC is supported by the U. S. Department of energy under contract
 No. W-7405-ENG-36.

\end{document}